\documentstyle[12pt]{article}
\newcounter{eqnn} \newcounter{eqs} \newcounter{secn}
\newcounter{subn}
\def\sec{\addtocounter{secn}{1}\setcounter{subn}{0}
\setcounter{eqs}{0}$\bf \thesecn $}
\def\sub{\addtocounter{subn}{1}  $\bf \thesecn.\thesubn $}
\def\eqn{\addtocounter{eqs}{1}\;\;\;(\thesecn.\theeqs)}
 
\def\lae{\;^{<}_{\sim} \;} \def\gae{\; ^{>}_{\sim} \;} 
\def\udd{u^{c}d^{c}d^{c}}  \def\ffr{\phi_{R}}
\def\sr{\psi_{R}} \def\ffi{\phi_{I}} \def\si{\psi_{I}}

\def\tt{\tilde{\theta}}

\begin{document}

\begin{titlepage}

\pagestyle{empty} \begin{center} {\Large \bf         
}\end{center}              \begin{flushright}  
LANCASTER-TH/9617\\              September
1996\end{flushright}              \vfill              
\begin{center} {\LARGE $\rm \udd$-Based Affleck-Dine
Baryogenesis}\\ \end{center} \vfill \begin{center} {\bf John
McDonald}\\              \vspace {0.1in}              School 
of Physics and Chemistry, \\  Lancaster University, \\  
Lancaster, \\ LA1 4YB\\United Kingdom 
\begin{footnote}{Address from 1st October 1996:  Dept. of
Physics, P.O.Box 9, University of Helsinki, FIN-00014 Helsinki,
Finland}\end{footnote} 
\begin{footnote}{e-mail: jmcd@laxa.lancs.ac.uk}\end{footnote}
\\ \vfill \end{center} \newpage
\begin{center} {\bf Abstract} \end{center}  

               We consider the possibility of a successful
Affleck-Dine mechanism along the                $\rm
u^{c}d^{c}d^{c}$ direction in R-parity symmetric extensions
of the minimal                supersymmetric Standard Model
(MSSM) which contain a gauge singlet superfield                
$\rm \phi$. Such gauge singlets commonly occur in extensions 
of the MSSM, for example in models which seek to account for                
neutrino masses. We consider a two scalar Affleck-Dine
mechanism,                with the flat direction stabilized 
by a non-renormalizible superpotential term of the form                   
$\rm \frac{\lambda}{M} \phi \udd \sim \frac{\lambda}{M}\phi
\psi^{3}$, where $\rm                \psi$ corresponds to
the gauge non-singlet flat direction.                We give 
approximate solutions of the scalar field equations of
motion which describe the                evolution of the
condensates and show that the final baryon asymmetry in this 
case is               suppressed relative to that expected
from the conventional single scalar Affleck-Dine mechanism,
based on a superpotential term of the form $\rm
\frac{\lambda}{4M} \psi^{4}$, by a factor $\rm
\left(\frac{m_{s}}{m_{\phi}+m_{s}}\right)^{1/2}$, where $\rm  
m_{s}$ is the soft supersymmetry breaking scalar mass and
$\rm m_{\phi}$ is the supersymmetric $\rm \phi$ mass.               
It is possible for the model to generate 
a baryon asymmetry even in the limit of 
unbroken B-L, so long as the gauge singlet condensate 
doesn't decay
until after anomalous electroweak B+L violation is out of
 equilibrium 
following the electroweak phase transition.
This condition is generally satisfied if 
all Dirac neutrino masses are less than around 10keV. 
This class of Affleck-Dine models can, in principle, be
 experimentally 
ruled out, for example by the observation of a Dirac mass for
 the 
$\rm \mu$ or $\rm \tau$ neutrino significantly
 larger than around 10keV together with a mostly Higgsino LSP. 
\end{titlepage}

{\bf \sec. Introduction}

          In supersymmetric (SUSY) models \cite{nilles}, the 
occurence of flat directions in the renormalizible scalar
potential of the minimal SUSY Standard Model (MSSM) and many 
of its extensions naturally leads to the possibility of
generating the baryon asymmetry of the Universe via the
decay of scalar field oscillations along such flat
directions. This possibility is the well-known Affleck-Dine
(A-D) mechanism for baryogenesis \cite{ad1}. Although in the 
limit of unbroken SUSY the renormalizible potential along
these flat directions is completely flat, once soft SUSY
breaking terms and non-renormalizible terms consistent with
the symmetries of the model are added there will be a
non-trivial potential. In the original A-D scenario
\cite{ad1} it was assumed that the
soft SUSY-breaking terms are the same as the zero
temperature SUSY-breaking terms, which are characterized by
a mass scale $\rm m_{s}$ of the order of 100GeV-1TeV 
\cite{nilles}. However, it has recently become clear that
the large energy density which exists in the early Universe
will also break SUSY, resulting in soft SUSY breaking terms
characterized by a mass scale typically of the order of the
Hubble parameter H \cite{Hb}. This large mass scale for the
SUSY breaking terms radically alters the evolution of the
scalar fields during and after inflation \cite{drt}. In the
original A-D mechanism, because the scalar field masses are 
much smaller than H during inflation, the classical scalar
fields are overdamped and effectively frozen in at their
initial values on horizon crossing, as generated by quantum 
fluctuations \cite{ad1,eu}. Therefore on the scale of the
observable Universe there is a large constant scalar field
over the whole Universe with an essentially random phase.
This then evolves into a coherently oscillating scalar
field, corresponding to a Bose condensate with a roughly
maximal asymmetry in the condensate particle number density. 
The subsequent decay of the condensate was shown to be
easily able to account for the baryon asymmetry of the
Universe \cite{ad1}. However, once mass terms of the order
of H for the scalar particles are introduced, this picture
completely changes \cite{drt}. Now the classical scalar
fields can evolve to the minimum of their potentials on a
time scale of the order of $\rm H^{-1}$. As a result, at the 
end of inflation, all the scalar fields will be at the
minimum of their potentials, with quantum fluctuations having 
an effect only on the scale of the horizon at the end of
inflation, which is much smaller than the scale of the
observable Universe. Therefore the baryon asymmetry coming
from the A-D condensate in this case will be determined
$dynamically$ by the evolution of the scalar fields during
the post-inflation era, with the scalar fields starting out
at the minimum of their potentials at the end of inflation.
Since the A-D mechanism is now dependent upon the details of 
the scalar potential, one has to consider each case
individually in order to determine the magnitude of the
resulting asymmetry. The asymmetry will be particularly
sensitive to which flat direction the scalar fields
oscillate along and to the form of the non-renormalizible
superpotential terms which determine the minimum of the
scalar potential and introduce the CP violation necessary in 
order to generate the baryon asymmetry \cite{drt}. 

                     In order to generate a baryon asymmetry 
from a condensate which decays prior to the electroweak phase 
transition (when anomalous $\rm B+L$ violation is in thermal 
equilibrium \cite{anom}) it is necessary for the condensate
to carry a non-zero B-L asymmetry. The lowest dimension
operators which characterize the B-L violating flat
directions in the MSSM are the dimension 2 (d=2) operator
$\rm LH_{u}$ and the d=3 operators $\rm u^{c}d^{c}d^{c}$,
$\rm d^{c}QL$ and $\rm e^{c}LL$ \cite{drt}. (These operators 
may be thought of as the superpotential terms responsible
for lifting the flat directions or as the scalar field
operators which are responsible for introducing explicit B-L 
violation into the scalar field equations of motion. These
are naturally connected by the relationship between the soft 
SUSY breaking terms and the superpotential terms
\cite{nilles,drt}). These operators characterize the flat
directions in the sense that the scalar field operator
characterizing a particular flat direction will have a
non-zero expectation value along that direction. We will
refer to the flat direction which gives a non-zero
expectation value to $\rm LH_{u}$ and to $\rm
\udd$ as the "$\rm LH_{u}$
direction" and the "$\rm \udd$ direction" respectively. (The A-D mechanism 
along the $\rm d^{c}QL$ and $\rm e^{c}LL$ directions will be 
essentially the same as that along the $\rm \udd$ direction;
we will concentrate on the $\rm \udd$ direction in the
following). The $\rm LH_{u}$ direction and the $\rm
u^{c}d^{c}d^{c}$ direction are orthogonal in the sense that
they cannot both be flat simultaneously \cite{drt}. The $\rm 
LH_{u}$ direction has recently been considered by a number
of authors \cite{drt,lhu}. In the present paper we will
focus on the $\rm \udd$ direction.

           The simplest implementation of the A-D mechanism
along the $\rm \udd$ direction would involve adding to the
MSSM superpotential a d=3 term of the form $\rm \udd$.
However, this term would be phenomenologically dangerous, as 
it would introduce large B violation into the MSSM unless
its coupling was extremely small. (For a review of the
constraints on B and L violating couplings in the MSSM see
reference \cite{bv}). For example, squark mediated proton
decay imposes the constraint $\rm |\lambda^{'}\lambda^{''}|
\lae 10^{-24}$  for the light quark generations, where $\rm
\lambda^{'}$ is the $\rm d^{c}QL$ coupling and $\rm
\lambda^{''}$ is the $\rm \udd$ coupling \cite{bv}. Such
dangerous B and L violating terms are usually eliminated
from the MSSM by imposing R-parity ($\rm R_{p}$)
\cite{nilles,bv}. Imposing $\rm R_{p}$ implies that the
first B-L violating operator in the MSSM which is nonzero
along the $\rm \udd$ direction is a dimension 6 operator,
$\rm u^{c}u^{c}d^{c}d^{c}d^{c}d^{c}$ \cite{drt}.  However,
as discussed in reference \cite{drt} (and briefly reviewed
in the present paper), in A-D models where the natural scale 
of the non-renormalizible terms is the Planck scale, d = 6
A-D models  can have an acceptably low B asymmetry only for
very low reheating temperatures $\rm T_{R}$, $\rm \theta
T_{R} \lae 10GeV$, where $\rm \theta$ is a CP violating
phase. d = 4 A-D models, on the other hand, can be
compatible with a much wider range of reheating
temperatures, up to $\rm 10^{9}GeV$ or more \cite{drt}. Thus 
with only the particle content of the MSSM, the $\rm \udd$
direction would be disfavoured in the simplest models (those 
based on Planck scale non-renormalizible terms) relative to
the $\rm LH_{u}$ direction, which can utilize the $\rm
R_{p}$-conserving d=4 operator $\rm (LH_{u})^{2}$. However,
if we were to consider extensions of the MSSM which involve
the addition of an $\rm R_{p}$-odd gauge singlet superfield
$\rm \phi$, then we could form the $\rm R_{p}$ conserving
d=4 operator $\rm \phi \udd$. The addition of such a gauge
singlet superfield to the MSSM is a very common and natural
feature of many extensions of the MSSM. In particular, in
models which seek to account for neutrino masses, the gauge
singlet superfield would correspond to a right-handed
neutrino superfield. It is the purpose of the present paper
to determine whether it is possible to generate the observed B
asymmetry along the $\rm \udd$ direction via the operator
$\rm \phi \udd$ and, if so, to compare the resulting asymmetry with
that coming from the more conventional $\rm LH_{u}$
direction.

                         The paper is organized as follows.
In section 2 we discuss the model and the minimization of
its scalar potential. In section 3 we consider the scalar
field equations of motion and the formation of the
coherently oscillating scalar field condensates. In section
4 we discuss the condensate particle asymmetries and the
resulting baryon asymmetry. In section 5 we discuss the
constraints on the reheating temperature after inflation. In 
section 6 we discuss the thermalization and decay of the
condensates and the upper limits on Dirac neutrino masses in
the limit of unbroken B-L. In section 7 we give our
conclusions.   \newpage {\bf \sec. d=4 Affleck-Dine
mechanism along the $\rm \udd$ direction}

            We will consider throughout the simplest scenario,
in which it is assumed that inflation occurs with an energy
density consistent with the density perturbations observed
by COBE, corresponding to $\rm H \approx 10^{14}GeV$
\cite{cobe}, with the inflaton $\rm \Phi$ subsequently undergoing
coherent oscillations about the minimum of its potential. We 
will also require that the reheating temperature $\rm T_{R}$ 
is low enough not to thermally regenerate gravitinos
\cite{tgrav},  which implies that $\rm T_{R}$ is less than
about $\rm 10^{10}GeV$, corresponding to H not much larger
than $\rm 1GeV$. After reheating we will assume that the
Universe is radiation dominated throughout, with no further
significant increase in entropy. In general, when inflation
ends H will not be much smaller that its value during
inflation (even in $\rm \Phi^{2}$ chaotic inflation, the
value of H when the inflaton $\rm \Phi$ starts oscillating
is not much smaller that $\rm 10^{13}GeV$ \cite{chi}).
Therefore the coherent oscillations of the A-D field, which
will begin once $\rm H \approx m_{s} \approx 100GeV$, will
begin during a matter dominated era, with the energy density 
of the Universe dominated by inflaton oscillations
\cite{drt}.

        It is now understood that in most supergravity
models, the energy density that exists in the early Universe 
will break SUSY, introducing soft SUSY breaking terms
characterized by a mass scale typically of the order of H
\cite{Hb}. We will therefore consider in the following soft
SUSY breaking terms of the form $${\rm V_{soft} = (m_{s}^{2} 
- c_{i}H^{2}) |\phi_{i}|^{2} + (B_{a}W_{2\;a}+h.c.) +
(A_{a}W_{n\;a}+h.c.)     \eqn},$$ where $\rm W_{2\;a}$ are
superpotential terms bilinear in the fields and $\rm
W_{n\;a}$ are terms of order n in the fields. $\rm A_{a}$
and $\rm B_{a}$ are defined by $\rm A_{a} = A_{a\;s}+
a_{a}H$ and $\rm B_{a} = B_{a\;s} + b_{a}H$, where $\rm
A_{a\;s}$ and $\rm B_{a\;s}$ may be thought of as the zero
temperature soft SUSY breaking terms from a hidden sector of 
N=1 supergravity \cite{nilles} whilst $\rm a_{a}H$ and $\rm
b_{a}H$ are due to SUSY breaking by the energy density in
the early Universe. We will assume throughout that  $\rm
A_{a\;s} \approx B_{a\;s} \approx m_{s}$. We will also
assume that $\rm a_{a}^{2} \approx b_{a}^{2} 
\approx |c_{a}| $. In most supergravity models we expect
that $\rm |c_{a}| \approx 1$ \cite{drt}, although in some
models $\rm |c_{a}|$ may be smaller; for example $\rm
|c_{a}| \approx 10^{-2}$ occurs in supergravity models with
a Heisenberg symmetry \cite{mur}. Then for the case with
$\rm c_{a} > 0$ the A-D scalar, corresponding to a
renormalizible flat direction in the scalar potential, will
have a non-zero value at the minimum of its scalar potential 
at the end of inflation, with the potential being stabilized 
by the contribution from the non-renormalizible terms in the
superpotential \cite{drt}. 

             To implement the d=4 A-D mechanism along the
$\rm \udd$ dirrection we will consider an $\rm R_{p}$
symmetric extension of the MSSM defined by the
superpotential $\rm W = W_{sm} + W^{'} + W_{\nu}$, where
$\rm W_{sm}$ is the MSSM superpotential, $\rm W^{'}$ is
defined by $${\rm W^{'} =  \frac{m_{\phi}}{2} \phi^{2} +
\frac{\lambda}{M} \phi u^{c}d^{c}d^{c}       + \frac{\eta}{4 
M} \phi^{4}   \eqn}$$ and $\rm W_{\nu}$ is given by $${\rm
W_{\nu} = \lambda_{\nu}\phi H_{u}L    \eqn}.$$ In addition
to these terms we would expect terms of the form $\rm \phi
d^{c}QL$ and $\rm \phi e^{c}LL$. For simplicity we will not
include these terms explicitly. The operator $\rm \udd
\equiv \epsilon_{\alpha \beta
\gamma}u^{c}_{\alpha}d^{c}_{\beta}d^{c}_{\gamma}$ (where
$\rm \alpha,\;\beta$ and $\rm \gamma$ are colour indices and 
generation indices are, for now, suppressed) is
antisymmetric in the $\rm d^{c}$ scalar fields. Therefore
the $\rm d^{c}$ should be from different generations, which
we will denote as $\rm d^{c}$ and $\rm d^{c\;'}$. With $\rm
u^{c}$, $\rm d^{c}$ and $\rm d^{c'}$ having different colour
indices, the F-term contribution to the scalar potential is
then given by $${\rm V_{F} = \sum \left| \frac{\partial 
W}{\partial \phi_{i}} \right|^{2} = m_{\phi}^{2}|\phi|^{2} +
\frac{|\lambda|^{2}}{M^{2}}| u^{c}d^{c}d^{c\;'}|^{2} +
\frac{|\eta|^{2}}{M^{2}}|\phi|^{6} }$$ $${\rm +
\frac{|\lambda|^{2}}{M^{2}}|\phi|^{2} \left[
|d^{c}d^{c\;'}|^{2} + |u^{c}d^{c}|^{2} + |u^{c}d^{c\;'}|^{2} 
\right]   }$$ $${\rm + \left[
m_{\phi}^{\dagger}\phi^{\dagger}(\frac{\lambda}{M}
u^{c}d^{c}d^{c\;'} + \frac{\eta}{M}\phi^{3})
+\frac{\lambda^{\dagger}\eta}{M^{2}}(u^{c}d^{c}d^{c'})^{
\dagger}\phi^{3} + h.c. \right] \eqn}.$$ The direction with only 
$\rm <u^{c}>$, $\rm <d^{c}>$, $\rm <d^{c\;'}>$ and $\rm
<\phi>$ non-zero is F-flat in the MSSM, with only the terms
in $\rm W^{'}$ lifting this flatness. The D-term
contribution to the scalar potential, $${\rm V_{D} = \sum
\frac{g_{i}^{2}}{2} | \Phi^{\dagger} T^{a}_{i} \Phi |^{2}
\eqn},$$ where the $\rm \Phi$ are the multiplets of the
gauge group i with generators $\rm T_{i}^{a}$, also vanishes 
so long as $\rm u^{c}$, $\rm d^{c}$ and $\rm d^{c\;'}$ have
different colour indices and $\rm |u^{c}|^{2} = |d^{c}|^{2}
= |d^{c\;'}|^{2} = v^{2}$. The phases of $\rm u^{c}$, $\rm
d^{c}$ and $\rm d^{c\;'}$ are not, however, fixed by the
MSSM F- and D-flatness conditions. We will see in the
following that the important phases for the A-D mechanism
are $\rm \delta_{v}$ and $\rm \delta_{\phi}$, where $\rm
<u^{c}d^{c}d^{c\;'}> = v^{3}e^{i \delta_{v}}$ and $\rm
<\phi> = v_{\phi}e^{i\delta_{\phi}}$.  \newpage {\bf \sub.
Potential minimization in the $\rm \eta \rightarrow 0$
limit}

       We first note that in the SUSY limit with H=0 and
with $\rm m_{\phi} \neq 0$,  there is a minimum with $\rm
v_{\phi} \neq 0$ which is degenerate with the $\rm v_{\phi}
= 0$ minimum and which could be phenomenologically
dangerous. To be precise, from equation (2.4) we see that
the SUSY minima correspond to $\rm \phi = 0$ and $\rm
\phi^{2} = -\frac{m_{\phi}M}{\eta}$ (with $\rm u^{c} = d^{c} 
= d^{c\;'} = 0$ at both minima). The $\rm \phi \neq 0$
minimum results in a dangerous $\rm \udd$ superpotential
term with coupling $\rm \frac{|\lambda|}{|\eta|^{1/2}}
\left(\frac{|m_{\phi}|}{M}\right)^{1/2}$. (In general we
would expect similar couplings for the $\rm d^{c}QL$ and
$\rm e^{c}LL$ terms). Thus, if we consider $\rm M \lae
M_{Pl}$ (where $\rm M_{Pl}$ is the Planck scale)  and $\rm
|m_{\phi}| \gae m_{s} \gae 10^{2}GeV$, then we see that $\rm
\frac{|\lambda|}{|\eta|^{1/2}} 
\left(\frac{|m_{\phi}|}{M}\right)^{1/2} \gae
\frac{|\lambda|}{|\eta|^{1/2}}10^{-8}$, which, for $\rm
|\lambda|$ and $\rm |\eta|$ not much smaller than 1, would
result in an unacceptable squark mediated proton decay rate
\cite{bv}. In order to avoid this danger we must therefore
ensure that there exists a minimum of the scalar potential
which has $\rm v \neq 0$ for large H but which evolves to
the $\rm v = v_{\phi} = 0$ minimum as H tends to zero. 

                With regard to the scale of the
non-renormalizible terms M, we will set this to equal the
Planck scale by convention. Then the coupling $\rm \lambda$
can take values small or large compared with 1, depending on 
the natural mass scale of the non-renormalizible terms.
Values much larger than 1 would correspond to the case where 
the natural mass scale of the non-renormalizible terms is
much smaller than the Planck scale, for example a grand
unification scale. On the other hand, if the natural mass
scale of the non-renormalizible terms was of the order of
the Planck scale, then we would expect that $\rm |\lambda|
\lae 1$. 

            Typically, we would not expect $\rm \lambda$ and 
$\rm \eta$ to be very different in magnitude. However, we
would like to be able to minimize the potential
analytically. We find that we can do this for the case of
$\rm |\eta|$ small compared with $\rm |\lambda|$ ($\rm
|\eta| \lae 0.1|\lambda|$ is sufficient), in which case it
may be shown that the terms in the scalar potential
proportional to $\rm \eta$ can, to a good approximation, be
neglected. We expect that the possibly more likely case with 
$\rm |\eta| \approx |\lambda|$ will be qualitatively
similar. In the following we will consider the minimization
of the potential in the $\rm \eta \rightarrow 0$ limit. 

              In the $\rm \eta \rightarrow 0$ limit the
scalar potential becomes $${\rm V = (|m_{\phi}|^{2} +
m_{s}^{2} - c_{\phi}H^{2}) v_{\phi}^{2}+
(m_{s}^{2}-c_{v}H^{2}) v^{2}  + 3
\frac{|\lambda|^{2}}{M^{2}} v_{\phi}^{2} v^{4} +
\frac{|\lambda|^{2}}{M^{2}} v^{6} + \left[ \frac{B_{\phi}
m_{\phi}}{2} v_{\phi}^{2} e^{2 i \delta_{\phi}}
\;\;+\;\;h.c.\right]  }$$ $${\rm + \left[
\frac{m_{\phi}^{\dagger}\lambda}{M}v_{\phi}v^{3}
e^{i(\delta_{v}-\delta_{\phi})}\;\;+\;\;h.c.\right] + \left[
\frac{A_{\lambda} \lambda}{M}v_{\phi}v^{3}e^{i(\delta_{v} +
\delta_{\phi})} \;\;+\;\;h.c.\right]    \eqn}.$$                 
By an choice of the phases of the scalar fields we can make
$\rm m_{\phi}$ and $\rm \lambda$ real. We may also choose
$\rm m_{\phi}$ to be positive. To a reasonable
approximation we can neglect the term proportional to $\rm
B_{\phi}$. This only contributes a term of the order of $\rm 
m_{\phi}(m_{s}+ a_{i}H)v_{\phi}^{2}$, which  is less than or 
of order of the $\rm (m_{\phi}^{2} + m_{s}^{2} -
c_{\phi}H^{2})v_{\phi}^{2}$ term. The phases $\rm
\delta_{\phi}$ and $\rm \delta_{v}$ will then adjust to
minimize the cross-terms (where we use "cross-terms" to
denote terms which are the sum of a term and its hermitian
conjugate). The potential will then have the form $${\rm V =
(m_{\phi}^{2} + m_{s}^{2} - c_{\phi}H^{2}) v_{\phi}^{2} +
(m_{s}^{2}-c_{v}H^{2}) v^{2}+  \frac{3 \lambda^{2}}{M^{2}}
v_{\phi}^{2} v^{4} + \frac{\lambda^{2}}{M^{2}} v^{6} - 2
\tilde{m}_{\phi} \frac{|\lambda|}{M} v_{\phi}v^{3}   
\eqn},$$ where 
$\rm \tilde{m}_{\phi} \equiv m_{\phi} +
|A_{\lambda}|$. In general, the minimum of this potential is
given by $${\rm v_{\phi} =  \frac{|\lambda|}{M} \frac{
\tilde{m}_{\phi}v^{3}}{(  (m_{\phi}^{2} + m_{s}^{2}-
c_{\phi} H^{2}) + \frac{3 \lambda^{2}}{M^{2}}v^{4})}   
\eqn}$$ and $${\rm v(v^{4}-
\frac{\tilde{m}_{\phi}M}{|\lambda|}v_{\phi}v + 2
v_{\phi}^{2}v^{2} + \frac{M^{2}}{3
\lambda^{2}}(m_{S}^{2}-c_{v}H^{2})  ) = 0   \eqn}.$$ We next 
consider how the minimum evolves from an initially large
value of $\rm c_{\phi}H^{2}$ . \newline {\bf (i) \underline{ 
$\rm c_{\phi}H^{2} > m_{\phi}^{2}+ m_{s}^{2}$ }} 

                   In general the minimum in this case is
given by $${\rm v^{4} \approx 
\frac{c_{\phi}}{\left(1-\alpha \right)} \frac{M^{2}H^{2}}{3
\lambda^{2}}  \eqn}$$ and $${\rm v_{\phi} \approx
\frac{1}{\sqrt{3}} \frac{\tilde{m}_{\phi}}{H}
\frac{(1-\alpha)^{1/2}}{c_{\phi}^{1/2}} \frac{v}{\alpha}
\approx \frac{1}{\sqrt{3}}
\frac{a_{\lambda}}{c_{\phi}^{1/2}}
\frac{(1-\alpha)^{1/2}}{\alpha} v  \eqn},$$ where $\rm
\alpha$ is the solution of $${\rm \alpha^{3} - (1
-\frac{a_{\lambda}^{2}}{c_{v}} - \frac{c_{\phi}}{c_{v}}  )
\alpha^{2}-\frac{5}{3} \frac{a_{\lambda}^{2}}{c_{v}} \alpha
+ \frac{2}{3} \frac{a_{\lambda}^{2}}{c_{v}} = 0    \eqn},$$
as may be seen by taking the $\rm H^{2}$ terms large
compared with the mass terms and substituting (2.10) into
(2.9) and (2.8). Typically $\rm (1-\alpha)$ is of the order
of 1. For example, if $\rm a_{\lambda}^{2}$ is small
compared with $\rm c_{v}$ then $\rm \alpha = (1 -
\frac{c_{\phi}}{c_{v}})$, whilst if $\rm a_{\lambda}^{2} =
c_{\phi} = c_{v}$ then $\rm \alpha = -2$. Therefore we can
roughly say, with $\rm c_{v} \approx c_{\phi} \approx
a_{\lambda}^{2}$, that $${\rm v \approx v_{\phi} \approx
c_{v}^{1/4}\frac{(MH)^{1/2}}{|\lambda|^{1/2}}   \eqn}.$$
Thus v and $\rm v_{\phi}$ are initially of the same order of 
magnitude. \newline {\bf (ii) \underline{ $\rm c_{\phi}H^{2} 
< m_{\phi}^{2}+ m_{s}^{2}$ }}      

          In this case we find that it is consistent to
assume that $\rm m_{\phi}^{2}+m_{s}^{2} \gae \frac{3
\lambda^{2}}{M^{2}} v^{4}$, in which case $\rm v_{\phi}$ is given
by $${\rm v_{\phi} \approx \frac{|\lambda|}{M}
\frac{\tilde{m}_{\phi}v^{3}}{\left( m_{\phi}^{2} + m_{s}^{2} 
\right)}    \eqn}.$$ Solving (2.9) for v, we find solutions $\rm
v_{\pm}$ given by $${\rm v_{\pm}^{4} =
-\frac{1}{4}\frac{M^{2}}{\lambda^{2}}
\frac{(m_{\phi}^{2}+m_{s}^{2})^{2}}{ \tilde{m}_{\phi}^{2}}
\left(1-
\frac{\tilde{m}_{\phi}^{2}}{(m_{\phi}^{2}+m_{s}^{2})}\right)
\left[ 1 \pm
\left[1-\frac{8}{3}\frac{\tilde{m}_{\phi}^{2}(m_{s}^{2}
-c_{v}H^{2})}{(m_{\phi}^{2}+m_{s}^{2}
-\tilde{m}_{\phi}^{2})^{2}}  \right]^{1/2}\right]   \eqn}.$$  
For $\rm c_{v}H^{2} >
m_{s}^{2}$, $\rm v_{-}^{4}$ is negative, and so there is a 
minimum at $\rm v_{+}$ with no barrier between $\rm v = 0$
and $\rm v = v_{+}$. Once $\rm c_{v}H^{2}$ is less than $\rm
m_{s}^{2}$, a barrier appears at $\rm v_{-}$. The minimum at 
$\rm v_{+}$ subsequently becomes unstable once  $${\rm
m_{s}^{2} - c_{v}H^{2} > \frac{3}{8} \frac{(2 |A_{\lambda}|
m_{\phi}+|A_{\lambda}|^{2} - m_{s}^{2}
)^{2}}{(m_{\phi}+|A_{\lambda}|)^{2}}    \eqn}.$$  For $\rm
m_{\phi}$ large compared with $\rm |A_{\lambda \;s}|$ and
$\rm m_{s}$ we see that we must have $\rm m_{s}^{2} >  
\frac{3}{2} |A_{\lambda\;s}|^{2}$ in order that the
dangerous $\rm v_{+} \neq 0$ minimum becomes destabilized as 
$\rm H \rightarrow 0$.  As $\rm m_{\phi} \rightarrow 0$,
this condition becomes $\rm m_{s}^{2} >  \frac{3}{8}
\frac{(|A_{\lambda\;s}|^{2}-m_{s}^{2})^{2}}{|A_{\lambda \;
s}|^{2}}$. Thus we see that it is non-trivial for the
dangerous $\rm v \neq 0$ minimum to become destabilized as
$\rm H \rightarrow 0$. So long as the potential $can$ be
destabilized, however, it will generally destabilize once $\rm
m_{s}^{2}$ is greater than $\rm c_{v} H^{2}$ up to a factor
of order 1. The values of the fields when the $\rm v_{+}$
minimum becomes unstable are then given by $${\rm v_{+}^{4}
\approx \frac{1}{4} \frac{M^{2}}{\lambda^{2}}
\left(\frac{m_{\phi}^{2}+m_{s}^{2}}{\tilde{m}_{\phi}^{2}}
\right) (\tilde{m}_{\phi}^{2}-m_{\phi}^{2}-m_{s}^{2})     
\eqn}$$ and  $${\rm v_{\phi} \approx \frac{1}{2}
\frac{(\tilde{m}_{\phi}^{2}-m_{\phi}^{2}
-m_{s}^{2})^{1/2}}{(m_{\phi}^{2}+m_{s}^{2})^{1/2}} v   
\eqn}.$$  Noting that $\rm |A_{\lambda}| \approx m_{s}$ for
$\rm c_{v} H^{2} \lae m_{s}^{2}$, we find that, for $\rm
m_{\phi} \gae m_{s}$, $${\rm v_{\phi} \approx  
\frac{1}{\sqrt{2}} 
\left(\frac{m_{s}}{m_{\phi}}\right)^{1/2} v    \eqn}, $$  
whilst for $\rm m_{\phi} < m_{s}$ $\rm v_{\phi} \approx v$.  
So for $\rm m_{\phi} > m_{s}$  we find that $\rm v_{\phi}$
becomes suppressed relative to v. \newpage{\bf \sec.  Bose
condensate formation}

            To discuss the formation of the Bose
condensates, that is to say, the way in which the scalar
fields start oscillating freely about the minimum of their
potentials, we consider the equations of motion of the
scalar fields. Strictly speaking we have four scalar fields;
$\rm u^{c}$, $\rm d^{c}$, $\rm d^{c\;'}$ and $\rm \phi$.
However, since we are considering the evolution of the
classical fields along a D-flat direction, we may impose
that the $\rm u^{c}$, $\rm d^{c}$ and $\rm d^{c\;'}$ fields
have the same magnitude. Although the phases of these 
fields could be different, the equations for $\rm u^{c}$,
$\rm d^{c}$ and $\rm d^{c\;'}$ are identical and the initial 
values of the fields are the same. Therefore we may assume
that their phases remain equal throughout. The equations of
motion are then given by, $${\rm \ddot{\psi_{R}} + 3H
\dot{\psi_{R}} =   -[(m_{s}^{2} - c_{v}H^{2}) \psi_{R} +
\frac{2 \lambda^{2}}{M^{2}}\phi_{R}^{2} \psi^{3}_{R} +
\frac{\lambda^{2}}{M^{2}} \psi_{R}^{5}    }$$ $${\rm  -
\beta \psi_{R}^{2}\phi_{R} - \alpha(t)
(c_{\theta}\phi_{R}\psi_{R}^{2}
+s_{\theta}(\psi_{R}^{2}\phi_{I}+2
\psi_{R}\phi_{R}\psi_{I})) ]     \eqn},$$  $${\rm
\ddot{\psi_{I}} + 3H \dot{\psi_{I}} =  -[(m_{s}^{2} -
c_{v}H^{2})\psi_{I} + \frac{2
\lambda^{2}}{M^{2}}\phi_{R}^{2} \psi^{2}_{R}\psi_{I} +
\frac{\lambda^{2}}{M^{2}} \psi_{R}^{4}\psi_{I}}$$ $${\rm -
\beta (\psi_{R}^{2}\phi_{I} -2\psi_{R}\psi_{I}\phi_{R}) -
\alpha(t) (s_{\theta}\phi_{R}\psi_{R}^{2} -
c_{\theta}(\psi_{R}^{2}\phi_{I}+2 \psi_{R}\phi_{R}\psi_{I})) 
]     \eqn},$$ $${\rm \ddot{\phi_{R}} + 3H \dot{\phi_{R}} =    
 -[(m_{\phi}^{2} + m_{s}^{2}- c_{\phi}H^{2})\phi_{R} - \beta 
\psi_{R}^{3} +\frac{3 \lambda^{2}}{M^{2}}
\psi_{R}^{4}\phi_{R} }$$ $${\rm - \gamma
(c_{\epsilon}\phi_{R}+s_{\epsilon}\phi_{I}) - \alpha
(c_{\theta}\psi_{R}^{3}+3 s_{\theta} \psi_{R}^{2}\psi_{I}) )  
]    \eqn},$$ and $${\rm \ddot{\phi_{I}} + 3H \dot{\phi_{I}} 
= -[(m_{\phi}^{2} + m_{s}^{2}- c_{\phi}H^{2})\phi_{I} - 3
\beta \psi_{R}^{2}\psi_{I} + \frac{3 \lambda^{2}}{M^{2}}
\psi_{R}^{4}\phi_{I} }$$ $${\rm - \gamma
(s_{\epsilon}\phi_{R}-c_{\epsilon}\phi_{I}) - \alpha
(s_{\theta}\psi_{R}^{3}-3 c_{\theta} \psi_{R}^{2}\psi_{I}) )  
]    \eqn},$$ where $\rm \psi$ represents the $\rm u^{c}$,
$\rm d^{c}$ and $\rm d^{c\;'}$ fields and we define 
$\rm \alpha$, $\rm \beta$ and $\rm \gamma$ by $\rm
\alpha = \left|\frac{A_{\lambda}\lambda}{M}\right|$, $\rm
\beta = \left|\frac{m_{\phi}\lambda}{M}\right|$ and $\rm
\gamma = \left|B_{\phi}m_{\phi}\right|$, where $\rm
\frac{m_{\phi}\lambda^{\dagger}}{M} = -\beta$ (we choose
this to be real and negative by an choice of phase), $\rm
(\frac{A_{\lambda}\lambda}{M})^{\dagger} = -\alpha
e^{i\theta}$ and $\rm  (B_{\phi} m_{\phi})^{\dagger} =
-\gamma e^{i \epsilon}$. 
$\rm c_{\theta}$ ($\rm c_{\epsilon}$) and $\rm s_{\theta}$ ($\rm 
s_{\epsilon}$) denote $\rm Cos\; \theta$ ($\rm Cos\; \epsilon$) and $\rm Sin\;  
\theta$ ($\rm Sin \;\epsilon$) respectively.
In writing these equations we have
assumed that the real parts of the fields are large compared 
with the imaginary parts, which turns out to be a reasonable 
approximation.   Throughout our discussion of the equations
of motion we will focus on the most likely form for the soft 
SUSY breaking terms in the early Universe, corresponding to
the case $\rm a_{i}^{2} \approx b_{i}^{2} \approx c_{i} \approx 1$
\cite{Hb,drt}. 

          Before the $\rm v \neq 0$ minimum is destabilized, 
the fields will be at the minimum of their potentials,
corresponding to setting the right hand side (RHS) of the
equations of motion to zero. The $\rm v \neq 0 $ minimum at
$\rm v = v_{+}$ will become destabilized once $\rm c_{v}
H^{2} \lae m_{s}^{2}$ and the $\rm \psi_{R}$ field will
begin to roll once $\rm H^{2} \lae m^{2}_{s}$. Once $\rm
\psi_{R}$ starts to roll, we will see that the other fields
$follow$ the minimum of their potentials as a function of
$\rm \psi_{R}(t)$ and oscillate about this minimum until
they become freely oscillating about the $\rm \psi_{R} = 0$
minimum of their potentials. 

              We first consider the evolution of the real
parts of the fields, beginning with $\rm \phi_{R}$. We 
consider the solution of the $\rm \phi_{R}$ equation of
motion in the limit where the terms proportional to $\rm
\psi_{R}^{4}\phi_{R}$, $\rm \gamma$ and $\rm s_{\theta}$ are 
neglected, which turns out to be a reasonable approximation.
 We will also treat $\rm \theta$ as a time-independent
constant. (We will comment on this later).  The $\rm
\phi_{R}$ equation of motion is then approximately given by
$${\rm \ddot{\phi_{R}} + 3H \dot{\phi_{R}} \approx   -[
(m_{\phi}^{2} + m_{S}^{2}- c_{\phi}H^{2})\phi_{R} -(\alpha
c_{\theta}+\beta)\psi_{R}^{3} ]      \eqn}.$$ From now on we
will neglect the $\rm c_{i}H^{2}$ mass terms in the equation 
of motion as these will quickly become negligible as the
Universe expands. We will also neglect the $\rm 3 H
\dot{\phi}$ and $\rm 3 H \dot{\psi}$ damping terms, since we
are considering $\rm m_{s} \gae H$. The effects of damping
will, however, be included in the time dependence of the
amplitude of oscillation of the fields. 

                      We wish to show that in the solution
of this equation $\rm \phi_{R}$ oscillates around the
minimum of its potential as a function of $\rm \psi_{R}$. To 
see this let $\rm \phi_{R} = \overline{\phi}_{R} + \delta
\phi_{R}$,  where $${\rm \overline{\phi}_{R} = \frac{(\alpha
c_{\theta}+\beta)}{m_{\phi}^{2}+m_{s}^{2}}\psi_{R}^{3}
\equiv \eta \psi_{R}^{3}    \eqn}$$ is the minimum of the
potential as a function of $\rm \psi_{R}(t)$. Then the $\rm
\delta \phi_{R}$ equation of motion is given by $${\rm
\delta \ddot{\phi}_{R} + 6 \eta \psi_{R} \dot{\psi}_{R}^{2}
\approx -(m_{\phi}^{2}+m_{s}^{2})\delta \phi_{R} + 3 \eta
m_{s}^{2} \psi_{R}^{3}      \eqn}.$$ In this we have used
$\rm \ddot{\sr} \approx -m_{s}^{2}\sr$, which will be shown
later to be true. $\rm \delta \ffr$ will then grow from an
initial value of zero until the mass term on the RHS of
(3.7) proportional to $\rm \delta \ffr$ becomes dominant,
after which $\rm \delta \ffr$ will oscillate about the
minimum $\rm \phi_{R} = \overline{\phi}_{R}$ with frequency
$\rm \approx (m_{\phi}^{2} + m_{s}^{2})^{1/2}$, the terms on 
the RHS proportional to $\rm \psi_{R}^{3}$ being rapidly
damped by the expansion of the Universe. The initial value
of the $\rm \delta \ffr$ oscillation amplitude is therefore
given by $\rm \delta \phi_{R\;o}$, where $${\rm \delta \phi_{R\;o} 
\approx \frac{3 \eta m_{s}^{2} }{m_{\phi}^{2} +
m_{s}^{2} }\psi_{R\;o}^{3}     \eqn}.$$
 It is straightforward to show that
$\rm \frac {\delta \phi_{R\;o} }{\overline{\phi}_{R\;o}}
\approx \frac{m_{s}^{2}}{m_{\phi}^{2}+m_{s}^{2}}$, which is
less than or about equal to 1. Eventually the amplitude of
$\rm \delta \ffr$ will become larger than that of $\rm
\overline{\phi}_{R}$, in which case $\rm \delta \ffr \approx 
\ffr$ will effectively oscillate freely around $\rm \ffr =
0$ with frequency $\rm (m_{\phi}^{2} + m_{s}^{2})^{1/2}$.

          We next consider the solution of the $\rm
\psi_{R}$ equation of motion. On introducing $\rm \phi_{R} = 
\delta \phi_{R}+ \overline{\phi}_{R}$, we find that, for
$\rm m_{s}^{2} \gae c_{v}H^{2}$, $\rm \psi_{R}$ begins
oscillating with a frequency approximately equal to $\rm
m_{s}$. This is not at first obvious since the $\rm
\psi_{R}^{2}\phi_{R}$ and $\rm \psi_{R}^{5}$ terms on the
RHS of the $\rm \psi_{R}$ equation of motion are initially
large ($\rm \sim (m_{s}m_{\phi})\psi_{R}$) compared with the 
$\rm m_{s}^{2}\psi_{R}$ term. However, it turns out that
there is a cancellation between these higher-order terms,
such that the sum of these terms on the RHS of (3.1)
contributes initially only $\rm \sim m_{s}^{2}\psi_{R}$ and 
then rapidly becomes small compared with the $\rm
m_{s}^{2}\psi_{R}$ term as $\rm \psi_{R}$ decreases with the 
expansion of the Universe. Therefore $\rm \psi_{R}$ will
essentially oscillate with frequency approximately equal to
$\rm m_{s}$ once the $\rm v = v_{+}$ minimum becomes
unstable.

              Thus we can summarize the evolution of the
real parts of the fields by  $${\rm \psi_{R}(t) \approx
A_{\psi}(t)Cos (m_{s}t)   \eqn}$$ and $${\rm \ffr(t) \equiv
\overline{\phi}_{R}  + \delta \phi_{R} \approx \eta 
\sr^{3}(t) + A_{\phi}(t) Sin \left((m_{\phi}+m_{s})t\right)  
\eqn},$$ where the time dependence of $\rm A_{\psi}(t)$ and
$\rm A_{\phi}(t)$ is due to the expansion of the Universe
during inflaton matter domination, $\rm A(t) \propto
a(t)^{-3/2}$, where $\rm a(t)$ is the scale factor. 

       We next consider the evolution of the imaginary parts 
of the fields. The evolution of $\rm \ffi$ and $\rm \si$ is
similar to the evolution of $\rm \ffr$ i.e. they follow the
minimum of their potentials as a function of $\rm \sr (t)$.
We first consider the solution of the $\rm \si$ equation of
motion.

      In the $\rm \si$ equation of motion we may roughly
absorb the terms proportional to $\rm \alpha c_{\theta}$
into  the terms proportional to $\rm \beta$. Thus we may
neglect these terms for now.  We will also set the $\rm
\beta \sr^{2}\phi_{I}$ term to zero for now. With these
assumptions only the phase $\rm \theta$ contributes to the 
imaginary parts of the fields. We will comment on these
assumptions later. 

          Suppose the $\rm \sr$ field starts oscillating at
$\rm t_{o}$. (For convenience we will set $\rm t_{o}$ equal
to 0 throughout). Initially, by a choice of the phase of the 
scalar fields, we can set the phase $\rm \theta$ to zero at
$\rm t_{o}$. The subsequent evolution of the phase $\rm
\theta(t)$ is then found  from $${\rm \alpha e^{i\theta(t)}
= -\left(\frac{(A_{\lambda \;s} +
a_{\lambda\;o}He^{i\sigma})\lambda}{M}\right)^{\dagger}(1
+a_{\lambda\;o}e^{i\sigma})^{-1} \eqn},$$ where $\rm \sigma$
is the phase difference between the $\rm A_{\lambda \; s}$
and $\rm a_{\lambda} \equiv a_{\lambda\;o}e^{i\sigma}$
terms. Since during matter domination $${\rm H =
\frac{H_{o}}{(1+\frac{3}{2}H_{o}t)}    \eqn}$$ where $\rm
H_{o} \equiv H(t_{o}) \approx m_{s}$, we see that   the
phase $\rm \theta(t)$ will reach its maximum 
roughly during the first $\rm 
\sr$ oscillation cycle, in a time $\rm \delta t \approx
H_{o}^{-1} \approx m_{s}^{-1}$. Since the condensates will
form during the first few oscillations of $\rm \psi_{R}$, it 
is a reasonable approximation to set $\rm \theta(t)$ to its
constant maximum value throughout.               With the
above assumptions the $\rm \si$ equation of motion can be
reasonably approximated by $${\rm \ddot{\si}  \approx -
\left[m_{s}^{2} \si + \frac{\lambda^{2}}{M^{2}}
\psi_{R}^{4}\psi_{I} -\alpha s_{\theta} \eta \psi_{R}^{5}
\right]     \eqn}.$$ From this we see that the minimum of
the $\rm \psi_{I}$ potential as a function of $\rm \psi_{R}$ 
is given by $${\rm \overline{\si} = \frac{\alpha
s_{\theta}\eta M^{2}}{\lambda^{2}
\left(1+\frac{m_{s}^{2}M^{2}}{\lambda^{2}\psi_{R}^{4}}\right
)} \sr      \eqn}.$$ Thus, noting that, at $\rm H \approx
m_{s}$, $\rm \frac{m_{s}^{2}M^{2}}{\lambda^{2}\psi_{R}^{4}}$ 
is small compared with 1, we see that $\rm \overline{\si}$
is initially proportional to $\rm \sr$. This will continue
until $\rm \sr$ decreases during its oscillation to the
point where $\rm
\frac{m_{s}^{2}M^{2}}{\lambda^{2}\psi_{R}^{4}} \gae 1$. 

               To understand the evolution of $\rm \psi_{I}$ 
let $\rm \si = \overline{\si} + \delta \si$. Substituting
into the $\rm \si$ equation of motion, we find that, for
$\rm \frac{m_{s}^{2}M^{2}}{\lambda^{2}\psi_{R}^{4}} \lae 1$,
 $\rm \delta \si$ satisfies
 $${\rm \delta \ddot{\psi}_{I}
\approx \frac{\alpha s_{\theta} \eta M^{2}}{\lambda^{2}}
m_{s}^{2}\sr - \frac{\lambda^{2}}{M^{2}}\psi_{R}^{4}\delta
\si  \eqn}.$$ Thus $\rm \delta \si $ will grow from $\rm
\delta \si =0$ to a value given by $${\rm \delta \psi_{I}
\approx \frac{\alpha s_{\theta}  \eta
m_{s}^{2}M^{4}}{\lambda^{4} \psi_{R}^{3}} \eqn}.$$ We see
that the condition $\rm
\frac{m_{s}^{2}M^{2}}{\lambda^{2}\psi_{R}^{4}} \lae 1$ is
equivalent to $\rm \delta \si \lae \overline{\psi}_{I}$. 
Initially $\rm \delta \si$ has a value $${\rm \delta \psi_{I\;o}
\approx \frac{s_{\theta} |A_{\lambda}|(|A_{\lambda}|
+m_{\phi})}{(m_{\phi}^{2}+m_{s}^{2})}  
\left(\frac{m_{s}}{m_{\phi} + m_{s}}\right) \sr  \eqn}.$$ 
Thus 
initially $\rm \frac{\delta \psi_{I\;o}}{\overline{\psi}_{I\;o}}
\approx \left(\frac{m_{s}}{m_{\phi} +
m_{s}}\right)^{2}s_{\theta}$, which is small compared with 1 
for $\rm m_{\phi}$ large compared with $\rm m_{s}$ or $\rm
\theta$ small compared with 1. During the subsequent $\rm \sr$ 
oscillation, we see that, so long as 
$\rm \frac{m_{s}^{2}M^{2}}{\lambda^{2}\psi_{R}^{4}} \lae 1$, 
$\rm \phi_{I}$ will be proportional to $\rm \psi_{R}$.
Therefore $\rm \si$ will initially be in phase with $\rm \sr$. 
However, as $\rm \psi_{R}$ decreases, for a period $\rm
\delta t$ during the $\rm \psi_{R}$ oscillation $\rm \delta
\psi_{I}$ will become larger than $\rm \overline{\psi}_{I}$
and the approximate equation of motion (3.15) will no longer 
be valid. During this time the $\rm m_{s}^{2} \psi_{I}$ term 
in the $\rm \psi_{I}$ equation of motion will dominate and
the $\rm \psi_{I}$ oscillation will continue with frequency
$\rm \approx m_{s}$. 
However, since during this period the effective
mass term in the $\rm \si$ equation of motion will differ
from $\rm m_{s}$ by a factor of order 1, it will be possible 
for the $\rm \psi_{I}$ phase to shift relative to $\rm
\psi_{R}$ by approximately $\rm m_{s}\delta t$. As a result, 
for a
 fraction 
$\rm m_{s} \delta t$ of the total $\rm \si$ oscillation,
there will be a phase shift approximately given by 
$\rm m_{s}\delta t$. 
Thus the average phase shift between $\rm \si$ and $\rm \sr$
over the period of the $\rm \si$ oscillation, which, 
as discussed in the next section, is
relevant for determining the $\rm \psi$ asymmetry, is given
by $\rm \delta_{p} \approx (m_{s} \delta t)^{2}$. $\rm
\delta t$ corresponds to the time during which $\rm
\psi_{R}^{4} \lae \frac{m_{s}^{2}M^{2}}{\lambda^{2}}$. For a 
matter dominated Universe, with $\rm \psi_{R} \propto
a(t)^{-3/2}$, we find that $\rm \delta_{p} \approx
(m_{s}\delta t)^{2} \approx
\left(\frac{m_{s}}{m_{\phi}+m_{s}}\right)^{1/2}
\left(\frac{m_{s}}{H}\right)^{2}$. $\rm \delta_{p}$ reaches
its largest value, $\rm \delta_{p} \approx 1$, once $\rm H
\approx
\left(\frac{m_{s}}{m_{\phi}+m_{s}}\right)^{1/4}m_{s}$.  We
also note that for $\rm \theta$ small compared with 1, $\rm
\frac{\si}{\sr} \approx \frac{\overline{\si}}{\sr} \approx 
\left(\frac{m_{s}}{m_{\phi} + m_{s}}\right)s_{\theta}$ is
small compared to 1, as has been assumed throughout.                    

       These results hold if (i) $\rm \alpha c_{\theta}$ is
small compared with $\rm \beta$ and (ii) if the $\rm \beta
\ffi \sr^{2}$ term in the $\rm \si$ equation of motion can
be neglected. The effect of the terms proportional to $\rm
\alpha c_{\theta}$ will be to effectively multiply the terms 
proportional to $\rm \beta$ by a factor $\rm \sim (1+
\frac{m_{s}}{m_{\phi}}c_{\theta})$. However, we can still
reasonably ignore the $\rm \beta \sr \si \ffr$ term in the
$\rm \si$ equation of motion, even with this factor. The
effect of including the $\rm \beta \ffi \sr^{2}$ term in the 
$\rm \si$ equation of motion (together with the factor from
(i)) turns out to be to approximately replace $\rm
s_{\theta}$ by $\rm s_{\tt}$, where $\rm s_{\tt} =
s_{\theta} + \left(\frac{m_{\phi}}{m_{\phi} + m_{s}}\right)
s_{\epsilon}$. Thus an imaginary part for $\rm \psi$ can
also be generated by the phase $\rm \epsilon$ so long as
$\rm \phi$ has a mass term in the superpotential.

          We finally consider $\rm \ffi$. With $\rm \ffi =
\overline{\ffi} + \delta \ffi$, where $${\rm \overline{\ffi} 
= \frac{(k \alpha s_{\theta} + \gamma \eta
s_{\epsilon})}{(m_{\phi}^{2}+m_{s}^{2})} \sr^{3}   \eqn}$$
and $${\rm k \approx
1+\frac{3(A_{\lambda}c_{\theta}+m_{\phi})}{(m_{s}^{2}
+m_{\phi}^{2})} m_{\phi}    \eqn},$$ the $\rm \ffi$ equation 
of motion can be written approximately as  $${\rm \delta
\ddot{\phi}_{I} + 6 \frac{(k \alpha s_{\theta} + \gamma \eta
s_{\epsilon})}{(m_{\phi}^{2}+m_{s}^{2})} \sr
\dot{\psi}^{2}_{R} \approx -(m_{\phi}^{2} + m_{s}^{2})\delta 
\ffi + 3 m_{s}^{2}\frac{(k \alpha s_{\theta} + \gamma \eta
s_{\epsilon})}{(m_{\phi}^{2}+m_{s}^{2})} \psi_{R}^{3}   
\eqn}.$$ 
Thus $\rm \delta \ffi$ will increase from zero to
an initial oscillation amplitude given by $${\rm \delta
\phi_{I\;o} \approx \frac{3m_{s}^{2}(k \alpha s_{\theta} +
\gamma \eta s_{\epsilon})}{(m_{\phi}^{2}+m_{s}^{2})^{2}}
\sr^{3}      \eqn}$$ and will subsequently oscillate freely
with frequency $\rm (m_{\phi}^{2}+m_{s}^{2})^{1/2}$. 
In general $\rm \delta
\phi_{I\;o} \lae \overline{\phi}_{I\;o}$. It is
straightforward to show that $\rm \delta \ffr$ and $\rm
\delta \ffi$ reach their maximum values and begin
oscillating in a time $\rm \delta t \sim
(m_{\phi}^{2}+m_{s}^{2})^{-1/2} \lae m_{s}^{-1}$ 
and so will become freely
oscillating within the first few oscillations of the $\rm
\psi_{R}$ field.

               It is important to emphasize that it is the
oscillations of $\rm \delta \ffr$ and $\rm \delta \ffi$
about the $\rm \overline{\phi}_{R}$ and $\rm
\overline{\phi}_{I}$ minima which evolve into the $\rm \phi$ 
Bose condensate as $\rm \overline{\phi}_{R}$ and $\rm
\overline{\phi}_{I}$ become smaller than $\rm \delta \ffr$
and $\rm \delta \ffi$. This occurs once $\rm H \lae
\left(\frac{m_{s}}{m_{\phi}+m_{s}}\right)m_{s}$. We will
show in the next section that the average $\rm \phi$
asymmetry over the $\rm \psi$ oscillation period $\rm m_{s}$ 
comes purely from $\rm \delta \phi$ and not from $\rm
\overline{\phi}$. 

              In summary, the $\rm \psi$ and $\rm \phi$
fields will begin to oscillate at  $\rm H \approx m_{s}$,
with the initial values of the oscillating scalar field
amplitudes relevant for the formation of condensate particle 
asymmetries given by   $${\rm \psi_{R\;o} \; \approx  v_{+} \;
\approx \frac{M^{1/2}}{|\lambda|^{1/2}} 
(m_{s}^{2}+m_{s}m_{\phi})^{1/4}      \eqn,}$$ $${\rm \psi_{I\;o}
\approx \left(\frac{\alpha s_{\tt} \eta
M^{2}}{\lambda^{2}}\right) \psi_{R\;o}  \approx s_{\tt}
\left(\frac{m_{s}}{m_{\phi}+m_{s}}\right) \psi_{R\;o}
\eqn,}$$  $${\rm \delta \phi_{R\;o} \approx \left(\frac{3 \eta 
m_{s}^{2}}{m_{\phi}^{2}+m_{s}^{2}}\right) \psi_{R\;o}^{3}
\approx \frac{m_{s}^{2}(m_{s}^{2}+m_{\phi}m_{s})^{1/2}}{
(m_{\phi}+m_{s})^{3}} \psi_{R\;o}    \eqn,}$$ and  $${\rm
\delta \phi_{I\;o} \approx \frac{(k \alpha s_{\theta} + \gamma
\eta s_{\epsilon})}{(m_{\phi}^{2}+m_{s}^{2})}
\psi_{R\;o}^{3}  \approx \frac{s_{\tt}
m_{s}^{3}(m_{s}^{2}+m_{\phi}m_{s})^{1/2}}{
(m_{s}^{2}+m_{\phi}^{2})^{2}} \psi_{R\;o}    \eqn,}$$   
where in the final expressions we have set $\rm
|A_{\lambda}| \approx |B_{\phi}| \approx m_{s}$ in order to
show the dependence on $\rm m_{s}$ and $\rm m_{\phi}$. 
However, as noted above and discussed further in the next
section, although the $\rm \psi$ scalar begins oscillating
at $\rm H \approx m_{s}$, for the case where $\rm m_{\phi}  
> m_{s}$ the full phase difference $\rm \delta_{p}$ between
$\rm \sr$ and $\rm \psi_{I}$ and the associated $\rm \psi$
particle asymmetry will only form once H is smaller than
$\rm \left(\frac{m_{s}}{m_{\phi}+m_{s}}\right)^{1/4}m_{s}$.

                      The approximations used in obtaining
these results are good for $\rm m_{\phi} \gg m_{s}$. For
$\rm m_{\phi} \lae m_{s}$ some of the assumptions made are
only marginally satisfied or even slightly violated
(although not strongly violated). However, we expect that
the above results for the initial amplitudes will still be
qualitatively correct, giving the correct order of magnitude 
for the resulting particle asymmetries. \newpage {\bf \sec.
Particle and Baryon Asymmetries.}

              In the limit where we retain only the mass
terms in the $\rm \psi$ and $\rm \phi$ equations of motion,
there is a global U(1) symmetry. This is broken by the B and 
L violating terms coming from the non-renormalizible terms
in the scalar potential, which give rise to a non-zero U(1)
charge in the condensate, corresponding to an asymmetry in
the number of $\rm \psi$ and $\rm \phi$ particles. We first
consider the $\rm \psi$ asymmetry. The asymmetry in the
number density of $\rm \psi$ particles is given by
$${\rm n_{\psi} = i(\psi^{\dagger}\frac{d\psi}{dt} -
\frac{d\psi}{dt}^{\dagger}\psi)  \equiv -2(\psi_{R}
\dot{\psi}_{I} - \psi_{I}\dot{\psi}_{R}) \eqn}.$$ 

     Note that in the case where $\rm \sr$ and $\rm \si$
oscillate with the same frequency, there must be a phase
difference between the oscillating $\rm \sr$ and $\rm \si$
fields in order to have a non-zero asymmetry. As discussed
in the last section, there is a time-dependent phase
difference $\rm \delta_{p}$ between $\rm \sr$ and $\rm \si$
(averaged over the period of oscillation $\rm m_{s}^{-1}$), which for the case 
$\rm \delta_{p} \lae 1$ is
given by $${\rm \delta_{p} \approx
\left(\frac{m_{s}}{m_{\phi}+m_{s}}\right)^{1/2}\left(\frac{m
_{s}}{H}\right)^{2}   \eqn}$$ 
and by $\rm \delta_{p} \approx 1$ otherwise. We can
characterize the $\rm \psi$ particle asymmetry by the
asymmetry at $\rm H \approx m_{s}$ that would evolve to the
correct $\rm \psi$ asymmetry at present, $${\rm n_{\psi}
\approx 2 \delta_{p} m_{\psi} \psi_{R\;o}\psi_{I\;o}    
\eqn},$$ where the value of $\rm \delta_{p}$ is determined
by the value of H at which the condensate thermalizes or
decays. 

                  For the case of the $\rm \phi$ asymmetry,
since, for $\rm m_{\phi} > m_{s}$, the $\rm \delta
\phi_{R,\;I}$ oscillate with a greater frequency than the $\rm
\overline{\phi}_{R,\;I}$, substituting $\rm\phi =
\overline{\phi} + \delta \phi$ into (4.1) shows that the
non-zero average asymmetry over the period $\rm H^{-1} \gae
m_{s}^{-1} \gae (m_{\phi}^{2}+m_{s}^{2})^{-1/2}$ will be that purely due to
$\rm \delta \phi_{R}$ and $\rm \delta \phi_{I}$. Thus we
find that the number density asymmetry in $\rm \phi$
particles is given by $${\rm n_{\phi} \approx
\left(\frac{m_{s}}{m_{\phi}+m_{s}}\right)^{5}
\frac{n_{\psi}}{\delta_{p}}     \eqn}.$$ The suppression of
$\rm n_{\phi}$ relative to $\rm n_{\psi}$ when $\rm m_{\phi} 
\gae m_{s}$ is significant, since in the limit where $\rm
m_{\phi} \rightarrow 0$ and $\rm \delta_{p} \rightarrow 1$ 
we may define an unbroken B-L asymmetry by assigning B=1 to
$\rm \phi$. This is broken by $\rm m_{\phi}$, suppressing
$\rm n_{\phi}$ and so preventing any possibility of a
cancellation between the B-L asymmetries coming from decay
of the $\rm \psi$ and $\rm \phi$ condensates. 

        We next compare the $\rm \psi$ asymmetry from the 
above $\rm \frac{\lambda}{M}\phi \psi^{3}$-type model with
that expected from the more conventional A-D mechanism based 
on a single field with a B violating term of the form $\rm  
\frac{\lambda}{4M}\psi^{4}$ together with the above form of H corrections 
to the SUSY breaking terms. In this case the initial values
of the $\rm \psi_{R}$ and $\rm \si$ fields are given by 
$${\rm \psi_{R\;o} \approx \left(\frac{1}{48}\right)^{1/4}
\frac{(m_{s}M)^{1/2}}{|\lambda|^{1/2}}      \eqn},$$ and
$${\rm \psi_{I\;o} \approx \frac{4 \alpha
s_{\theta}\psi_{R}^{3}}{m_{s}^{2}}    \eqn},$$ with $\rm
\sr$ and $\rm \si$ subsequently oscillating with a phase
difference of the order of 1. Thus in this case we find that
$${\rm n_{\psi} \approx \frac{\alpha s_{\theta} 
M}{|\lambda|} \psi_{R\;o}^{2}     \eqn},$$  with the
asymmetry being fully formed at $\rm H \approx m_{s}$.
Therefore, for the case where the CP violating phases $\rm
s_{\theta}$ and $\rm s_{\tt}$ and the coupling $\rm \lambda$ 
have the same values in both cases,  we find that the
asymmetry from the $\rm \frac{\lambda}{M}\phi \psi^{3}$-type 
superpotential term is related to that from the $\rm
\frac{\lambda}{4M}\psi^{4}$ term by  $${\rm n_{\psi} \approx 
\delta_{p} \left(\frac{m_{s}}{m_{\phi}+m_{s}}\right)^{1/2}
n_{\psi\;o}     \eqn},$$ where $\rm n_{\psi\;o}$ is the
asymmetry expected from the conventional single field A-D
mechanism. Thus we see that, even with $\rm \delta_{p}
\approx 1$, for $\rm m_{\phi} > m_{s}$ there is a
suppression of the $\rm \psi$ asymmetry by a factor $\rm
\left(\frac{m_{s}}{m_{\phi}+m_{s}}\right)^{1/2}$ relative to 
that expected from the conventional single-field A-D
mechanism.   

              The baryon asymmetry from the $\rm \psi$
condensate is simply the asymmetry $\rm n_{\psi}$ multiplied
by a factor for the number of baryons produced per $\rm
\psi$ decay. Since we have in fact three condensates, 
corresponding to $\rm u^{c}$, $\rm d^{c}$ and $\rm
d^{c\;'}$, each of which carries baryon number 1/3, we see
that $\rm n_{B} = n_{\psi}$. The baryon asymmetry to entropy 
ratio after reheating is then found by noting that the ratio 
of the $\rm \psi$ number asymmetry to the inflaton energy
density during inflaton oscillation domination is constant.
Since the reheating temperature is given by $\rm \rho_{I} =
k_{T}T_{R}^{4}$ and the entropy density by $\rm s =
k_{s}T_{R}^{3}$ (with $\rm k_{T} = \frac{\pi^{2}g(T)}{30}$
and $\rm k_{s} = \frac{2\pi^{2}g(T)}{45}$, where $\rm g(T) = 
g_{b}(T) + \frac{7}{8}g_{f}(T)$ and $\rm g_{b}(T)(g_{f}(T))$
are the number of bosonic (fermionic) degrees of freedom in
thermal equilibrium at temperature T \cite{eu}), it follows
that the baryon-to-entropy ratio is given by $${\rm
\frac{n_{B}}{s} = 
\frac{k_{s}}{k_{T}T_{R}}\frac{n_{B}}{\rho_{I}}   \eqn}$$
Thus, with the energy density dominated by inflaton
oscillations when the Bose condensate forms at $\rm H
\approx m_{s}$ and with $\rm n_{B} = n_{\psi} \approx 
\frac{s_{\tt}\delta_{p} m_{s}^{2}M}{|\lambda|}
\left(\frac{m_{s}}{m_{\phi}+m_{s}}\right)^{1/2}$, we find
that $${\rm \frac{n_{B}}{\rho_{I}} \approx \frac{8 \pi}{3}
\frac{s_{\tt}\delta_{p}}{|\lambda| M_{Pl}}
\left(\frac{m_{s}}{m_{s} +m_{\phi}}\right)^{1/2}    \eqn},$$ 
where we have used $\rm M \equiv M_{Pl}$ and $\rm
|A_{\lambda}| \approx m_{s}$. Therefore $${\rm
\frac{n_{B}}{s} \approx   \frac{2 \pi
s_{\tt}\delta_{p}}{|\lambda|} \left(\frac{m_{s}}{m_{\phi}
+m_{s}}\right)^{1/2}\frac{T_{R}}{M_{Pl}}    \eqn}.$$

                Comparing this with the observed asymmetry,
$\rm \frac{n_{B}}{s} \approx 10^{-10}$, and noting that the
thermal gravitino regeneration constraint implies that $\rm
\frac{T_{R}}{M_{Pl}} \lae 10^{-9}$ \cite{tgrav}, we see
that $\rm m_{\phi}$ cannot be too large compared with $\rm
m_{s}$ if $\rm \frac{s_{\tt}\delta_{p}}{|\lambda|}$ is not
large compared with 1, as we would expect if the natural
scale of the non-renormalizible terms was less than or of
the order of $\rm M_{Pl}$. On the other hand, even if $\rm
\frac{s_{\tt}\delta_{p}}{|\lambda|}$ is large compared
with 1, the suppression factor $\rm
\left(\frac{m_{s}}{m_{\phi}+m_{s}}\right)^{1/2}$ would still 
allow $\rm T_{R}$ to be consistent with the observed B
asymmetry for values of $\rm T_{R}$ up to the gravitino
constraint, so long as $\rm m_{\phi}$ was sufficiently
large.

             Thus we can conclude that, in the $\rm R_{p}$
symmetric MSSM extended by the addition of a gauge singlet
scalar, it is indeed possible to have a successful d=4
two-scalar $\rm \phi \psi^{3}$-type A-D mechanism along the
$\rm \udd$ direction. The resulting baryon asymmetry,
assuming that the asymmetries are able to fully form and do
not thermalize or decay before the phase $\rm \delta_{p}(t)$ 
has reached its maximum value (we discuss this possibility
in the next section), receives an overall suppression by a
factor approximately $\rm \left(\frac{m_{s}}{m_{\phi}+
m_{s}}\right)^{1/2}$ relative to that expected in the case
of a conventional d=4 $\rm \psi^{4}$-type A-D mechanism
based on a single A-D scalar field. This suppression factor
allows for a wider range of non-renormalizible couplings and 
reheating temperatures to be compatible with the observed
baryon asymmetry than in the case of the conventional A-D
mechanism. \newpage {\bf \sec. No-Evaporation Constraint}

                An important constraint on the reheating
temperature comes from the requirement that the interaction 
of the condensate scalars with the radiation energy density
due to inflaton decays prior to reheating does not lead to
the condensate thermalizing  via scattering with the
background plasma before the particle asymmetries can be
established \cite{drt}. We refer to this as the
no-evaporation constraint. 

                    We first consider the $\rm \psi$
condensate. In general, there are two ways in which the
condensate can be destroyed at a given value of H;
thermalization and decay. Thermalization of the condensate
will occur if (i) the rate of scattering of the thermal
plasma particles from the condensate scalars, $\rm
\Gamma_{s}$, is greater than H
and (ii) if the mass of the initial and final state 
particles are such that the scattering process is kinematically
allowed and the scattering particles in the plasma
are not Boltzmann suppressed. Decay of the condensate via
tree level two-body decays will occur if (i) the decay rate
of the condensate scalars in their rest frame, $\rm
\Gamma_{d}$, is greater than H and (ii) if the final state
particles, of mass $\rm \lambda_{\psi}<\psi>$, where 
$\rm \lambda_{\psi}$ is a gauge or Yukawa coupling and 
$\rm <\psi>$ is the amplitude of the $\rm \psi$ oscillation, are
lighter than the condensate scalars. Since the time over
which the real and imaginary parts of the fields start
rolling and so the asymmetries start to develop is of the
order of $\rm H^{-1}$ at $\rm H \approx m_{s}$, we must
ensure that thermalization and decay does not occur on a
time scale small compared with $\rm m_{s}^{-1}$.

               Prior to reheating, the radiation energy
density coming from inflaton decays during inflaton matter
domination  corresponds to a background plasma of particles
at a temperature $\rm T_{r}$, where, assuming that the decay
products thermalize, $\rm T_{r}$ is given by \cite{drt,eu}
  $${\rm T_{r} \approx
k_{r}(M_{Pl}HT_{R}^{2})^{1/4}       \eqn}, $$ where $\rm
k_{r} = \left(\frac{3}{50\pi k_{T}}\right)^{1/8} \approx
0.4$ (using $\rm g(T) \approx 100$). 

For the case of the $\rm \psi$ condensate, the particle
asymmetry begins to form at $\rm H \approx m_{s}$, at which
time $\rm \delta_{p}(t) \approx
\left(\frac{m_{s}}{m_{s}+m_{\phi}}\right)^{1/2}
\left(\frac{m_{s}}{H}\right)^{2} \approx
\left(\frac{m_{s}}{m_{s}+m_{\phi}}\right)^{1/2}$, and
subsequently grows until $\rm \delta_{p} \approx 1 $ at $\rm 
H \approx
\left(\frac{m_{s}}{m_{\phi}+m_{s}}\right)^{1/4}m_{s}$. 

             Thermalization of 
the $\rm \psi$ condensate is possible if 
the rate of scattering of the plasma particles in equilibrium at
temperature $\rm T_{r}$ from the
condensate scalars is sufficiently large. For the
case of t-channel scattering of condensate scalars 
from plasma
 fermions via SU(3) gauge boson or Yukawa fermion exchange 
interactions, the scattering rate is given by
$\rm \Gamma_{s} \approx k_{\Gamma}\sigma
\lambda_{\psi}^{4}T_{r}$, where $\rm \lambda_{\psi}$
is the gauge or Yukawa coupling, 
$\rm \sigma = \frac{1}{x(1+2x)}$ for the gauge boson exchange
and $\rm \sigma = \frac{1}{4}log\left(\frac{1}{x}\right)$ for the
Yukawa fermion exchange, 
with $\rm x = \frac{m_{A}^{2}}{9T^{2}}$
and $\rm m_{A}$ the mass of the exchanged 
gauge boson or fermion,
and where for scattering from a
 single Dirac fermion $\rm k_{\Gamma}\approx
\frac{1}{12\pi^{3}} \approx 3x10^{-3}$. The
condition for the plasma to be able to thermalize the
condensate is then that $\rm \Gamma_{s} \gae H$, which
implies that 
$${\rm \lambda_{\psi} \gae 5x10^{-2}
\left(\frac{H}{m_{s}}\right)^{3/16}
\left(\frac{10^{10}GeV}{T_{R}} \right)^{1/8}
\left(\frac{1}{\sigma}\right)^{1/4}
\eqn}. $$ 
We see that this will be satisfied by the gauge couplings, the 
top quark Yukawa coupling and, marginally, the bottom
 quark Yukawa coupling. 
Thus, in order to prevent the
 thermalization of the condensate, we must
require that $\rm 
\lambda_{\psi}<\psi> \gae 3T$ for these couplings, 
to ensure that the associated scattering processes
are kinematically suppressed. To be safe, we will conservatively
require that 
\newline $\rm \lambda_{\psi}<\psi> \gae 30T$, in order to 
suppress scattering from plasma particles with energy larger than
the mean thermal energy 3T.
For the case
 of the $\rm \udd$ direction, we note that the smallest
Yukawa coupling to the condensate scalar $\rm \psi$,
which will typically involve a linear combination
of all three generations of down squark, will
equal the b quark Yukawa coupling up to a factor of the order of
1. Thus, in general,
kinematically suppressing the b quark Yukawa interaction
will ensure that the $\rm \psi$ condensate is not thermalized. 
Although the b quark Yukawa coupling,
$\rm \lambda_{b} \approx 4x10^{-2}$ for 
the MSSM with equal Higgs doublet expectation values,
only marginally 
satisfies the condition for thermalization, and so
might allow the condensate to form without 
large suppression of the asymmetry,
if we were to consider t-channel fermion exchange 
scattering from light gauginos and squarks 
in the plasma, then the constraint (5.2)
 would apply to the combination 
$\rm \lambda_{\psi} \approx (g\lambda_{b})^{1/2}$, 
where g is a gauge coupling.
 This would exceed the lower bound (5.2). 
Thus we will conservatively assume that 
the b quark Yukawa interaction must be kinematically suppressed
in order to avoid thermalization
 of the condensate on a time scale 
small compared with $\rm H^{-1}$ and
 so to allow the asymmetries to form.
 
The condition
$\rm \lambda_{\psi}<\psi> \gae \kappa T_{r}$ 
corresponds to  $${\rm T_{R} 
\lae \frac{\lambda_{\psi}^{2}}{\kappa^{2} k_{r}^{2} \lambda}
((m_{\phi}+m_{s})M_{Pl})^{1/2}
\left(\frac{H}{m_{s}}\right)^{3/2}    \eqn},$$ where during
matter domination, $${\rm <\psi> =
\left(\frac{a(t)}{a_{o}}\right)^{3/2}\psi_{R\;o} \approx 
\frac{M^{1/2}\left(m_{s}^{2}
+m_{s}m_{\phi}\right)^{1/4}}{\lambda^{1/2}}\left(\frac{H}{m_
{s}}\right)    \eqn},$$ with $\rm a_{o}$ the scale factor at 
$\rm H \approx m_{s}$. Using equation (4.11) together with
the observed B asymmetry, $\rm \frac{n_{B}}{s} \approx
10^{-10}$, we find that $${\rm T_{R} \lae
\frac{10^{-5}M^{3/4}m_{s}^{1/4}}{\sqrt{2 \pi s_{\tt}
\delta_{p}}} \frac{\lambda_{\psi}}{\kappa k_{r}}
\left(\frac{m_{\phi}+m_{s}}{m_{s}}\right)^{1/2}
\left(\frac{H}{m_{s}}\right)^{3/4} \eqn}.$$  Thus, with $\rm
\lambda_{\psi} \equiv \lambda_{b} \approx 4x10^{-2}$
for the b quark Yukawa coupling and with $\rm \kappa = 30$, 
we find that the condensate will survive if 
$${\rm T_{R} \lae \frac{1}{\sqrt{s_{\tt}}} 
\left(\frac{m_{\phi}+m_{s}}{m_{s}}\right)^{3/4}
\left(\frac{H}{m_{s}}\right)^{7/4} 10^{7}GeV    \eqn},$$ 
where we have used $\rm m_{s} \approx 10^{2}GeV$.  The form
of this constraint depends on whether we impose the
no-evaporation constraint before the asymmetry has fully
formed or not. If we consider the constraint to apply at
$\rm H \approx m_{s}$, when the $\rm \psi$ asymmetry is
minimal, then the $\rm T_{R}$ upper bound is proportional to 
$\rm \left(\frac{m_{\phi}+m_{s}}{m_{s}} \right)^{3/4}$. On
the other hand, if we allow the $\rm \psi$ asymmetry to grow 
to its maximum value before thermalization, then the $\rm
T_{R}$ upper bound is proportional to $\rm
\left(\frac{m_{\phi}+m_{s}}{m_{s}}\right)^{5/16}$.   In both 
cases the upper bound on $\rm T_{R}$ is weakened by having
$\rm m_{\phi} > m_{s}$. We see that, with $\rm m_{\phi} 
\lae m_{s}$, $\rm T_{R}$ can take values up to around $\rm
\frac{10^{7}GeV}{\sqrt{s_{\tt}}}$
 without preventing the formation
of the asymmetry. 
With $\rm m_{\phi} > m_{s}$
this constraint becomes weaker, allowing $any$
reheating temperature up to the thermal gravitino limit to
be compatible with the no-evaporation constraint, regardless 
of the value of $\rm \tt$, so long as $\rm m_{\phi}$ is
 sufficiently 
large. 

          We also have to check that the $\rm \psi$
condensate does not decay before the condensates fully form
at $\rm H \approx \left(
\frac{m_{s}}{m_{\phi}+m_{s}}\right)^{1/4}m_{s}$. If $\rm
\lambda_{\psi}<\psi> \gae m_{s}$, then the two-body
tree-level decay will be kinematically suppressed. This 
occurs if $${\rm \frac{\lambda_{\psi}}{\sqrt{\lambda}} \gae
\left(\frac{m_{s}}{M_{Pl}}\right)^{1/2}   \eqn},$$ which
will almost certainly be satisfied. The higher-order decay modes 
will then be suppressed by a factor of at least $\rm
\left(\frac{m_{s}}{\lambda_{\psi}<\psi>}\right)^{4}$, which
gives $\rm \Gamma_{d} \ll H$ at $\rm H \approx m_{s}$. 
Thus $\rm \psi$ decay is ineffective at $\rm H \approx m_{s}$ and 
the no-evaporation constraint is the correct condition
for the initial $\rm \psi $ asymmetry to be able to form. 

            A second, perhaps less important, constraint on
the reheating temperature comes from the requirement that
$\rm \frac{n_{B}}{s} \approx 10^{-10}$ can be consistent
with non-renormalizible operators whose natural mass scale
is the Planck scale, corresponding to $\rm |\lambda| \lae
1$. In fact, from (4.11), we see that, with $\rm
\frac{n_{B}}{s} \approx 10^{-10}$, the reheating temperature 
is given by  $${\rm T_{R} \approx \frac{|\lambda|}{2\pi
s_{\tt}\delta_{p}}
\left(\frac{m_{\phi}+m_{s}}{m_{s}}\right)^{1/2} 10^{9}GeV    
\eqn}.$$ (The conventional $\rm \psi^{4}$ d=4 models give
the same result but with $\rm m_{\phi} \rightarrow 0$ and
$\rm \delta_{p} \approx 1$ \cite{drt}). Thus we see that
$\rm |\lambda| \lae 1$ is necessary in order for $\rm T_{R}$ 
to be consistent with the thermal gravitino bound. Although
$\rm |\lambda| \lae 1$ is possible even if the mass scale of 
the non-renormalizible operators is small compared with $\rm 
M_{Pl}$, it is most natural for the case of Planck scale
operators. Thus d=4 models are most naturally consistent
with the thermal gravitino bound when the mass scale of the
non-renormalizible operators corresponds to the Planck mass. 
For the case of a conventional single field A-D mechanism
with d=6 operators, the reheating temperature is given by
\cite{drt} $${\rm T_{R} \approx
\left(\frac{|\lambda|^{1/2}}{s_{\theta}}\right) 10GeV
\eqn}.$$  Thus in this case the reheating temperature is
expected to be very low compared with the thermal gravitino
bound for the case of Planck scale non-renormalizible
operators, although for non-renormalizible operators with a
smaller mass scale, which would naturally have $\rm
|\lambda| \gg 1$, larger reheating temperatures would be
possible. Therefore we see that d=4 models are favoured for
the case of Planck scale operators, naturally allowing for a 
much wider range of reheating temperatures than d=6 models,
whilst d=6 models are favoured if the natural mass scale of
the non-renormalizible operators is much smaller than the
Planck scale.
 \newpage{\bf \sec. Condensate Thermalization and
Decay after Reheating and an Upper Limit on Dirac
Neutrino Masses.}

     After the inflaton decays, the Universe will be
radiation dominated with $\rm H =
\frac{k_{H}T^{2}}{M_{Pl}}$,  where $\rm k_{H} =
\left(\frac{4 \pi^{3} g(T)}{45}\right)^{1/2} \approx 17$. We 
first show that $\rm \psi$ will typically thermalize at a
temperature large compared with the temperature of the
electroweak phase transition $\rm T_{EW} \approx 10^{2}GeV$. 
$\rm \Gamma_{s} \gae H$ occurs if $\rm \lambda_{\psi} \gae
\left(\frac{k_{H}}{k_{\Gamma}}\right)^{1/4}\left(\frac{T}{
M_{Pl}}\right)^{1/4}
\left(\frac{1}{\sigma}\right)^{1/4}$.
Since $\rm T_{R} \lae 10^{10}GeV$, this will be satisfied if 
$\rm \lambda_{\psi} \gae 5x10^{-2}
\left(\frac{1}{\sigma}\right)^{1/4}$.
Thus so long as the scattering process in not kinematically or
Boltzmann suppressed, corresponding to
 $\rm \frac{\lambda_{\psi} <\psi>}{3T}$ smaller than
1, the condensate will thermalize.  With, for $\rm T <
T_{R}$,  $${\rm <\psi> \approx \frac{
10^{-2}}{\lambda^{1/2}}\left(\frac{T_{R}}{10^{10}GeV}\right)
^{1/2} \left(\frac{m_{\phi}+m_{s}}{m_{s}}\right)^{1/4}
\left(\frac{T}{m_{s}}\right)^{3/2}  \eqn},$$ where we are
using $\rm m_{s} \approx 10^{2}GeV$ throughout, $\rm  
\frac{\lambda_{\psi} <\psi>}{T} \lae 1$ requires that $${\rm
\frac{\lambda_{\psi}}{\sqrt{\lambda}} \lae 10^{4}
\left(\frac{m_{s}}{m_{\phi}+m_{s}}\right)^{1/4}
\left(\frac{10^{10}GeV}{T_{R}}\right)^{1/2}
\left(\frac{m_{s}}{T}\right)^{1/2}   \eqn}. $$ This will
typically be satisfied for some value of T larger than $\rm
T_{EW}$. 

   For the case of the gauge singlet $\rm \phi$ condensate,
we first note that 
since the $\rm \phi$ scalar is $\rm R_{p}$ odd, 
it can decay only if it is not the 
lightest supersymmetric partner (LSP). 
The most rapid possible $\rm \phi$ decay will 
correspond to a tree-level two-body 
decay to a left-handed neutrino 
and a neutralino via the neutrino Yukawa 
coupling $\rm \lambda_{\nu}$. This will be kinematically  
allowed so long as one
 of the neutralinos has a mass less than the 
$\rm \phi$ scalar mass. In particular, this will occur if one of 
the neutralinos is the LSP, as is strongly favoured 
by the possibility of neutralino cold dark matter. 
The $\rm \phi$ decay rate will then depend on the proportion of 
light mass eigenstate neutralino(s) contained in the weak eigenstate 
fermion in $\rm H_{u}$. This will in turn depend on whether the  
light neutralino in question is mostly gaugino or Higgsino. 
If it is mostly Higgsino, then
the decay will occur via the neutrino
 Yukawa coupling with coupling strength 
approximately equal to $\rm \lambda_{\nu}$.
On the other hand, if it is
 mostly gaugino, then we would expect the 
coupling of the 
$\rm \phi$ scalar to the light neutralino to have an
 addtitional suppression factor 
of the order of $\rm \frac{m_{W}}{\mu}$, where $\rm \mu$ 
corresponds to the $\rm \mu H_{u}H_{d}$
term in the MSSM superpotential \cite{nilles}. 
Typically, for $\rm \mu \lae 1TeV$, 
this factor will not be much smaller than about 0.1.

   Thus in the case of a mostly Higgsino 
light neutralino (or, more generally, for the 
case where the mass term $\rm \mu$
 is small compared with the $\rm 
\phi$ scalar mass), tree-level two-body 
$\rm \phi$ decay will occur via the neutrino 
Yukawa coupling $\rm \lambda_{\nu}$ if $\rm
\Gamma_{d} \approx
 \frac{\alpha_{\nu}}{4}(m_{\phi}^{2}+m_{s}^{2})^{1/2}
\gae H$,
where $\rm \alpha_{\nu} \approx
\frac{\lambda_{\nu}^{2}}{4\pi}$. This is satisfied if $${\rm
\lambda_{\nu} \gae 
\left(\frac{16\pi k_{H}}{\left(m_{\phi}+m_{s}\right)M_{Pl}}
\right)^{1/2}  T  \eqn}.$$ 
The decay is kinematically allowed so long as $\rm
\lambda_{\nu}<\phi> \lae (m_{\phi}^{2}+m_{s}^{2})^{1/2}$. With,
 for $\rm T
\lae T_{R}$, $${\rm <\phi> \approx
\frac{T_{R}^{2}}{m_{s}M_{Pl}}\frac{m_{s}^{2}(m_{s}^{2}+m_{s}
m_{\phi})^{3/4}}{ (m_{\phi}+m_{s})^{3}}
\left(\frac{M_{Pl}}{\lambda}\right)^{1/2}
\left(\frac{T}{T_{R}}\right)^{3/2}    \eqn},$$ this
condition requires that $${\rm
\frac{\lambda_{\nu}}{\sqrt{\lambda}} \lae
10^{4}\left(\frac{10^{10}GeV}{T_{R}}\right)^{1/2}\left(
\frac{m_{\phi}+m_{s}}{m_{s}}\right)^{13/4}
\left(\frac{m_{s}}{T}\right)^{3/2}   \eqn}.$$
It is straightforward to show that, in the cases of most interest to us here, 
the decay of the $\rm \phi$ condensate typically occurs before 
it can thermalize by scattering. The condition for 
the condensate to thermalize by scattering via the Yukawa 
coupling $\rm \lambda_{\nu}$
is that $\rm \Gamma_{s} \approx k_{\Gamma}\sigma
\lambda_{\nu}^{4}T \gae H = \frac{k_{H}T^{2}}{M_{Pl}}$, where 
$\rm \sigma \approx
 \frac{1}{2}log\left(\frac{3T}{\lambda_{\psi}<\psi>}\right)$. 
This implies that
$${\rm \lambda_{\nu} \gae 5x10^{-4}
\left(\frac{T}{m_{s}}\right)^{1/4} 
\left(\frac{1}{\sigma}\right)^{1/4}
   \eqn}.$$
This lower bound is typically larger than the
lower bound on $\rm \lambda_{\nu}$ coming from $\rm \phi$ decay.
It is possible that thermalization could occur by scattering from
light 
sleptons and SU(2) gauginos in the plasma, which would replace
$\rm \lambda_{\nu}$ by $\rm (\lambda_{\nu}g)^{1/2}$,
where $\rm g \approx 0.6$ is the SU(2) gauge coupling. 
However,
 the lower bound on $\rm \lambda_{\nu}$ will still typically
be larger than that coming from $\rm \phi$ decay.
In particular, this is
 true for the important case of $\rm \phi$ decay 
below the temperature of the electroweak phase transition, which we 
discuss below. 
Another possibility for 
thermalizing the condensate is via inverse decays and
related $\rm 2 \rightarrow 1$ processes,             
which are expected to have a rate 
$\rm \Gamma_{inv} \approx \kappa \lambda_{\nu}^{2} T$, 
where $\rm \kappa \ll 1$. Although
 at high temperature this rate can be large
compared with the $\rm \phi$ decay rate, for temperatures at or below 
the electroweak phase  
transition temperature, which are of most interest to us here, 
the direct decay rate will be much larger than 
the rate of thermalization 
via inverse decays. 
In all this we have assumed 
that the $\rm \phi$ decay occurs via the neutrino 
Yukawa coupling. It is also possible that the $\rm \phi$ 
condensate could 
decay via the non-renormalizible superpotential coupling 
$\rm \frac{\lambda}{M} \phi \udd$ once $\rm <\psi>$ is introduced, 
which gives an effective $\rm \phi \psi \psi$ coupling. However, 
it is straightforward to check 
that this effective coupling is in general
much smaller than the typical 
values of $\rm \lambda_{\nu}$ considered in neutrino
mass models, and so may be neglected when discussing $\rm \phi$ 
condensate decay.

     Thus we see that the $\rm \phi$ condensate can 
evade decay until $\rm T < T_{EW}$ if
$\rm \lambda_{\nu}$ is sufficiently small.
To see what
 this implies for the baryon asymmetry and for neutrino
masses, we first note that the gauge singlet scalar $\rm
\phi$ will typically correspond to a linear combination of
the three right-hand sneutrino generations. Thus $\rm
\lambda_{\nu}$ will typically correspond to the largest
neutrino Yukawa coupling up 
to a factor of around $\rm \frac{1}{\sqrt{3}}$.
In the $\rm m_{\phi} \rightarrow 0$ limit,
corresponding to the case where the neutrinos have Dirac
masses,  we can define an $unbroken$ B-L asymmetry by
defining $\rm \phi$ to have B=1. However, so long as the
$\rm \phi$ condensate decays $after$ the electroweak phase
transition has occured, the effect of anomalous
electroweak B+L violation \cite{anom}, which is in thermal
equilibrium at temperatures larger than $\rm T_{EW}$, will
be to alter only the B asymmetry coming from the thermalized 
$\rm \psi$ condensate and so prevent a cancellation of the B 
asymmetry coming from the $\rm \psi$ and $\rm \phi$
conensate, even though the net B-L asymmetry will be zero.
Thus so long as the $\rm \phi$ condensate decays at $\rm
T_{d\;\phi} \lae T_{EW}$ in the limit $\rm m_{\phi}
\rightarrow 0$, there will still be a B asymmetry. The
magnitude of the B asymmetry will be essentially the same as 
that previously calculated from the decay of the $\rm \psi$
condensate alone, since the magnitude of the net B asymmetry 
will, up to a factor of the order of 1,
equal that of the B-L asymmetry coming from
$\rm \psi$ thermalization at $\rm T
 \gae T_{EW}$. The coupling $\rm
\lambda_{\nu}$ is related to the heaviest neutrino mass in
the $\rm m_{\phi} \rightarrow 0$ limit by $\rm
m_{\nu}
\approx 10^{2}\lambda_{\nu}GeV$. From this and (6.3) we find 
that, for the case of a mostly Higgsino light neutralino, 
the $\rm \phi$ condensate decays at a temperature $\rm
T_{d\;\phi} \approx 10^{7}m_{\nu}$.  Thus the condition $\rm 
T_{d\;\phi} \lae T_{EW}$ implies that all Dirac neutrino
masses should satisfy $${\rm m_{\nu} \lae 10keV   \eqn},$$  
assuming that $\rm \phi$ corresponds to a roughly equal
combination of the three sneutrino generations, as we would
generally expect. 
This is true for the case of a light neutralino which is 
mostly Higgsino, or more generally for the case 
where the $\rm \phi$ decay via the 
Yukawa coupling $\rm \lambda_{\nu}$ 
is completely unsuppressed. From this we see that so long as the 
Dirac neutrino masses are all below about 10keV, 
the $\rm \phi$ condensate will generally decay 
below the temperature of the electroweak phase transition and 
so a baryon asymmetry will be generated even in 
the limit of B-L conservation.   
On the other hand, for the case of, for example,
a mostly gaugino LSP 
this upper bound could be increased to around 
100keV or more, depending on the particular gaugino LSP mass 
eigenstate and the $\rm \mu$ parameter. 
These upper bounds should be compared with the present
experimental upper bounds on the neutrino masses, $\rm
m_{\nu_{\tau}} < 24MeV$, $\rm m_{\nu_{\mu}} < 160keV$ and
$\rm m_{\nu_{e}} < 5.1eV$ \cite{numass}. From these we see
that the requirement that a non-zero B asymmetry can be
generated in the limit of unbroken B-L in the case where 
the LSP is a neutralino imposes a non-trivial upper
bound on $\rm m_{\nu_{\mu}}$ and $\rm m_{\nu_{\tau}}$. 
In particular, we see that it would be possible, in
principle, to experimentally rule out this class of
Affleck-Dine models, for example if neutrinos with Dirac masses 
significantly larger
than around 10keV were found to exist together with an 
LSP which was mostly Higgsino. 
We also note that an unbroken 
B-L asymmetry would rule out the possibility of the d=4 $\rm 
LH_{u}$ direction, leaving the d=4 $\rm \udd$ direction as
the unique d=4 possibility in the case of B-L conserving
models with non-zero neutrino masses. 

                   For the case with $\rm m_{\phi} \neq 0$,
the neutrinos will gain Majorana masses via the see-saw
mechanism \cite{seesaw}, with $\rm m_{\nu} \approx
\frac{(\lambda_{\nu}100GeV)^{2}}{m_{\phi}}$. Thus in this
case $\rm T_{d\;\phi}$ is given by $${\rm T_{d\;\phi}
\approx 10^{4} \left(\frac{m_{\nu}}{1eV}\right)^{1/2}
\left(\frac{m_{\phi}}{100GeV}\right)^{1/2}
\left(\frac{m_{\phi}+m_{s}}{100GeV}\right)^{1/2}GeV     
\eqn}.$$ Therefore typically the $\rm \phi$ condensate will
decay at $\rm T > T_{EW}$ in this case. 

     Throughout the above discussion we have assumed that
the Universe is radiation dominated. It is straightforward
to show that this is indeed the case. The $\rm \psi$
condensate would dominate the energy density only once T
satisfies $${\rm T \lae
\frac{(m_{s}^{2}+m_{s}m_{\phi})^{1/2}}{\lambda}
\frac{T_{R}}{M_{Pl}}   \eqn},$$ which is typically satisfied 
only for temperatures less than around $\rm 10^{-7}GeV$. For the
$\rm \phi$ condensate the energy density is even less than
the $\rm \psi$ condensate, by a factor $\rm
\left(\frac{m_{s}}{m_{\phi}+m_{s}}\right)^{3}$. Thus 
the Universe will be radiation dominated when the
condensates thermalize or decay.   \newpage {\bf \sec.
Conclusions}

                          We have considered the possibility 
of generating the observed baryon asymmetry via an
Affleck-Dine mechanism based on the renormalizible F- and
D-flat $\rm \udd$ direction of the SUSY Standard Model. In
order to avoid breaking $\rm R_{p}$ whilst allowing a d=4
superpotential term to lift the flatness and drive
baryogenesis, we considered extensions of the SUSY Standard
Model which have additional gauge singlets $\rm \phi$, such
as commonly occur in models which seek to account for
neutrino masses. In such models the $\rm \udd$ direction
becomes potentially as important as the more commonly
considered $\rm LH_{u}$ direction. We have shown that the
A-D mechanism based on the d=4 operator $\rm \phi \psi^{3}$, 
where $\rm \psi$ is a gauge non-singlet A-D field, can
indeed (for an appropriate choice of parameters) generate
the baryon asymmetry whilst allowing the scalar fields to
evolve to a phenomenologically acceptable minimum. The
resulting asymmetry is suppressed relative to the asymmetry
coming from the more conventional $\rm \psi^{4}$-based A-D
mechanism (such as the $\rm LH_{u}$ direction) by a factor
$\rm \left(\frac{m_{s}}{m_{\phi}+m_{s}} \right)^{1/2}$, all
couplings and CP phases being taken equal, where $\rm
m_{\phi}$ is the SUSY $\rm \phi$ mass term and $\rm m_{s}$
is the soft SUSY breaking mass scale. This suggests that
$\rm m_{\phi}$ cannot be much larger than $\rm m_{s}$, if
the observed baryon asymmetry is to be generated without
requiring very small couplings in the non-renormalizible
terms. The requirement that the initial condensate particle 
asymmetry can form before the 
condensate is thermalized imposes an upper bound on the 
reheating temperature
of $\rm 
\frac{10^{7}GeV}{\sqrt{\theta}}$ in the limit where $\rm m_{\phi}
\lae m_{s}$, where $\rm \theta$
 is the CP violating phase responsible for the
baryon asymmetry. The upper bound becomes weaker if $\rm m_{\phi}  
\gae m_{s}$. Thus with a sufficiently
large $\rm m_{\phi}$ or small $\rm \theta$ the whole 
range of reheating temperatures up to the thermal gravitino
constraint 
can be compatible with the initial formation of an asymmetry.
The $\rm \psi$ condensate will typically thermalize
before the electroweak phase transition occurs. Then in the
limit of unbroken B-L (for which case the $\rm \udd$
direction is the unique d=4 possibility), which corresponds
to $\rm m_{\phi} \rightarrow 0$, a B asymmetry can be
generated only if the $\rm \phi$ condensate 
decays below the temperature of the electroweak phase
transition, when anomalous electroweak B+L violation is out
of thermal equilibrium. This 
will generally be true if all neutrino masses
are less than around 10keV. 
In the case where the LSP is a neutralino, 
or more generally where there is a neutralino mass eigenstate 
lighter than the $\rm \phi$ scalar, 
the $\rm \phi$ decay condition imposes a non-trivial upper
limit on Dirac neutrino masses. For example, for the case 
of a mostly Higgsino LSP, the upper bound is around 
$\rm 10keV$, whilst for a 
mostly gaugino LSP this bound could increase to 
around 100keV or more, depending on the $\rm \mu$ 
parameter of the MSSM and the 
particular gaugino LSP mass eigenstate.  Thus
the observation of a Dirac mass for the $\rm \mu$ or $\rm
\tau$ neutrino significantly larger than 
10keV together with a mostly Higgsino 
LSP, for example, would experimentally
rule out this class of Affleck-Dine models.  \newline
\newline \newline The author would like to thank Kari Enqvist for 
useful comments. This research was supported by the PPARC.

\end{document}